\documentclass[a4paper,final]{revtex4}
\usepackage[latin1]{inputenc}
\usepackage[T1]{fontenc}
\usepackage[english]{babel}
\usepackage[draft]{graphicx}
\usepackage{amsmath}
\usepackage{amssymb}
\usepackage{mathrsfs}
\usepackage{textcomp}
\usepackage{tabularx}
\usepackage{color}
\usepackage{array}
\usepackage{float}

\begin{document}

\title{0-$\pi$ quantum transition in a carbon nanotube Josephson junction: Universal phase dependence and orbital degeneracy}

\author{R. Delagrange}
\affiliation{Laboratoire de Physique des Solides, CNRS, Univ. Paris-Sud, Universit\'e Paris Saclay, 91405 Orsay Cedex, France.}
\author{R. Weil}
\affiliation{Laboratoire de Physique des Solides, CNRS, Univ. Paris-Sud, Universit\'e Paris Saclay, 91405 Orsay Cedex, France.}
\author{A. Kasumov}
\affiliation{Laboratoire de Physique des Solides, CNRS, Univ. Paris-Sud, Universit\'e Paris Saclay, 91405 Orsay Cedex, France.}
\author{M. Ferrier}
\affiliation{Laboratoire de Physique des Solides, CNRS, Univ. Paris-Sud, Universit\'e Paris Saclay, 91405 Orsay Cedex, France.}
\author{H. Bouchiat}
\affiliation{Laboratoire de Physique des Solides, CNRS, Univ. Paris-Sud, Universit\'e Paris Saclay, 91405 Orsay Cedex, France.}
\author{R. Deblock \footnote{author to whom correspondence should be addressed : deblock@lps.u-psud.fr}}
\affiliation{Laboratoire de Physique des Solides, CNRS, Univ. Paris-Sud, Universit\'e Paris Saclay, 91405 Orsay Cedex, France.}

\begin{abstract}
In a quantum dot hybrid superconducting junction, the behavior of the supercurrent is dominated by Coulomb blockade physics, which determines the magnetic state of the dot. In particular, in a single level quantum dot singly occupied, the sign of the supercurrent can be reversed, giving rise to a $\pi$-junction. This 0-$\pi$ transition, corresponding to a singlet-doublet transition, is then driven by the gate voltage or by the superconducting phase in the  case of strong competition between the superconducting proximity effect and Kondo correlations. In a two-level quantum dot, such as a clean carbon nanotube, 0-$\pi$ transitions exist as well but, because more cotunneling processes are allowed, are not necessarily associated to a magnetic state transition of the dot.
In this proceeding, after a review of 0-$\pi$ transitions in Josephson junctions, we present measurements of current-phase relation in a clean carbon nanotube quantum dot, in the single and two-level regimes. In the single level  regime, close to orbital degeneracy and in a regime of strong competition between local electronic correlations and superconducting proximity effect, we find that the phase diagram of the phase-dependent  transition is a universal characteristic of a discontinuous level-crossing quantum transition at zero temperature. In the case where the two levels are involved, the nanotube Josephson current exhibits a continuous 0-$\pi$ transition, independent of the superconducting phase, revealing a different physical mechanism of the transition.

\end{abstract}

\maketitle

\section{Introduction}

Josephson junctions \cite{Josephson1962} refer to any non superconducting material sandwiched between two superconductors. The simplest Josephson junction (JJ), an insulator between two superconductors, is passed through by a non dissipative current $I=I_C \sin(\varphi)$, $I_C$ being the critical current and $\varphi$ the phase difference of the superconducting order parameters. When a normal metal is inserted between the superconductors, the transmission of Cooper pairs takes place through Andreev Bound States (ABS), that are confined in the normal region at an energy below the gap \cite{Nazarov2009}. Due to the boundary conditions, the energy of these bound states depends on $\varphi$, leading to a phase dependence of the supercurrent: the current-phase relation (CPR).

In a single level QD-JJ, the physics of the ABS is governed by four characteristic energies: the coupling $\Gamma=\Gamma_L+\Gamma_R$ ($\Gamma_L$ and $\Gamma_R$ are the coupling respectively to the left and right reservoirs, $\Gamma_L/\Gamma_R$ is the asymmetry), the charging energy $U$, the level energy in the dot $\epsilon_d$ and the superconducting gap of the contacts $\Delta$. We focus in this article on the intermediate regime $\Gamma\approx U\approx\Delta$, where the Coulomb Blockade is strong enough to impose a well defined occupancy and the coupling sufficient to observe a supercurrent \cite{defranceschi2010}.  The transfer of Cooper pairs then involves cotunneling processes, strongly dependent on the dot's occupancy. When this occupancy is even, one has a 0-junction whose amplitude follows the transmission of the dot. For an odd occupancy, the first non-zero contribution to the supercurrent involves fourth order processes, that imply reversing the spin-ordering of the Cooper pair. The sign of the supercurrent is thus reversed and its amplitude strongly reduced, this is called a $\pi$-junction. Experimentally, the supercurrent can be precisely changed by tuning the parity of the dot with a gate voltage \cite{Vandam2006,Cleuziou2006,Jorgensen2007}. 
 
 In addition, an oddly occupied dot gives rise to Kondo effect. The interaction of the local magnetic moment with delocalized conduction electrons through spin flip processes leads, in the normal state, to the formation of a strongly correlated state. This Kondo singlet state is characterized by the screening of the dot's magnetic moment and by a resonance in the density of states at the Fermi energy for temperatures below the Kondo temperature $T_K$ \cite{Pustilnik2004,Goldhaber-Gordon1998,Cronenwett1998}. In the superconducting state, when $k_B T_K < \Delta$, the Kondo screening is destroyed by superconducting correlations and does not affect the $\pi$-junction. But for $k_B T_K\gg\Delta$, the unpaired electron's spin is involved in a Kondo singlet that opens a well transmitted channel in the system and facilitates the transfer of Cooper pairs. Therefore the 0-junction is recovered and the supercurrent is enhanced due to the cooperation between superconductivity and Kondo effect \cite{Clerk2000,Glazman1989,Jorgensen2007,Eichler2009}. Since the Kondo temperature depends on $U$, $\Gamma$ and $\epsilon$ \cite{Haldane1978}, the junction can be tuned from 0 to $\pi$ by varying these parameters for a fixed parity and value of $\Delta$ \cite{Maurand2012}.
Measuring the CPR of the single-level QD Josephson junction directly gives insights into the magnetic state of the system: a doublet if one measures a $\pi$-junction, a singlet state (purely BCS or Kondo) otherwise. Here, we are particularly interested in the specific regime of the 0-$\pi$ transition, where the system undergoes a level-crossing quantum transition. The fundamental state of the system, 0 or $\pi$, depends on the superconducting phase $\varphi$, meaning that the magnetic state of the system (singlet or doublet) can be controlled by this parameter \cite{Vecino2003,choi2004,Siano2004,Bauer2007,Karrasch2008,Meng2009,Luitz2010,Luitz2012}. The gate and/or phase induced 0-$\pi$ transition measured on the CPR is directly related to the behaviour of the Andreev states \cite{Eichler2007,grove-rasmussen_superconductivity-enhanced_2009,Pillet2010,Deacon2010,Chang2013,Kim2013,Lee2014}, which coincide, for a certain range of parameters, with the Yu-Shiba-Ruzinov states when the quantum dot is oddly occupied \cite{kirsanskas_yu-shiba-rusinov_2015,jellinggaard_tuning_2016}. 

In a multi-level quantum dot, the picture described previously is not valid anymore, as predicted theoretically \cite{Rozhkov2001,Zazunov2010,Lee2010,Karrasch2011,Droste2012}. The measurement of the current-phase relation is no longer a good indicator of the effective magnetic moment of the dot. Indeed, as soon as several energy levels participate in the transport, the available cotunelling processes are different and the properties of the wave-functions become determinant, making 0 and $\pi$-junction possible both for even and odd occupations. This multi-level effects on the supercurrent in a quantum dot based Josephson junction have been experimentally observed by van Dam et al. \cite{Vandam2006} with an InAs nanowire. But no phase dependence of the supercurrent was shown and the analysis of the experiment was complicated by the fact that, unlike in carbon nanotube, the exact electronic configuration is not known.  

The aim of this work is to present the investigation of the CPR in a clean carbon nanotube (CNT) QD, where the orbital levels are nearly degenerate. Our results show that distinct behaviors emerge depending on the number of levels involved in transport. This number is determined by the occupancy and the relative widths of the nearly degenerate orbital levels. In our sample, for most filling factors, the system is well understood in a single-level description. In this regime, for odd electronic occupation and intermediate transmission of the contacts,  we probe experimentally the existence of this phase-driven 0-$\pi$ transition and demonstrate its universal character.  On the other hand, in some odd diamonds with nearly degenerate orbital levels, we qualitatively confirm theoretical predictions about the gate dependence of the supercurrent in the two-level regime and its high sensitivity to the precise configuration of the two orbital states involved in transport. In addition, the phase dependence of the supercurrent shows a continuous 0-$\pi$ transition with a complete cancellation of the amplitude of the Josephson current, in contrast with the first order single-level 0-$\pi$ transition. 

The paper is organized as follows. In a first part, we remind the reader of the basics of the physics involved and review the different kinds of existing 0-$\pi$ transitions, and their relation with quantum phase transitions. Then, we present our experimental results.

\section{Josephson effect in a carbon nanotube quantum dot and 0-$\pi$ transitions: State of the art}

\begin{figure}[htbp]
\begin{center}
\includegraphics[height=4cm]{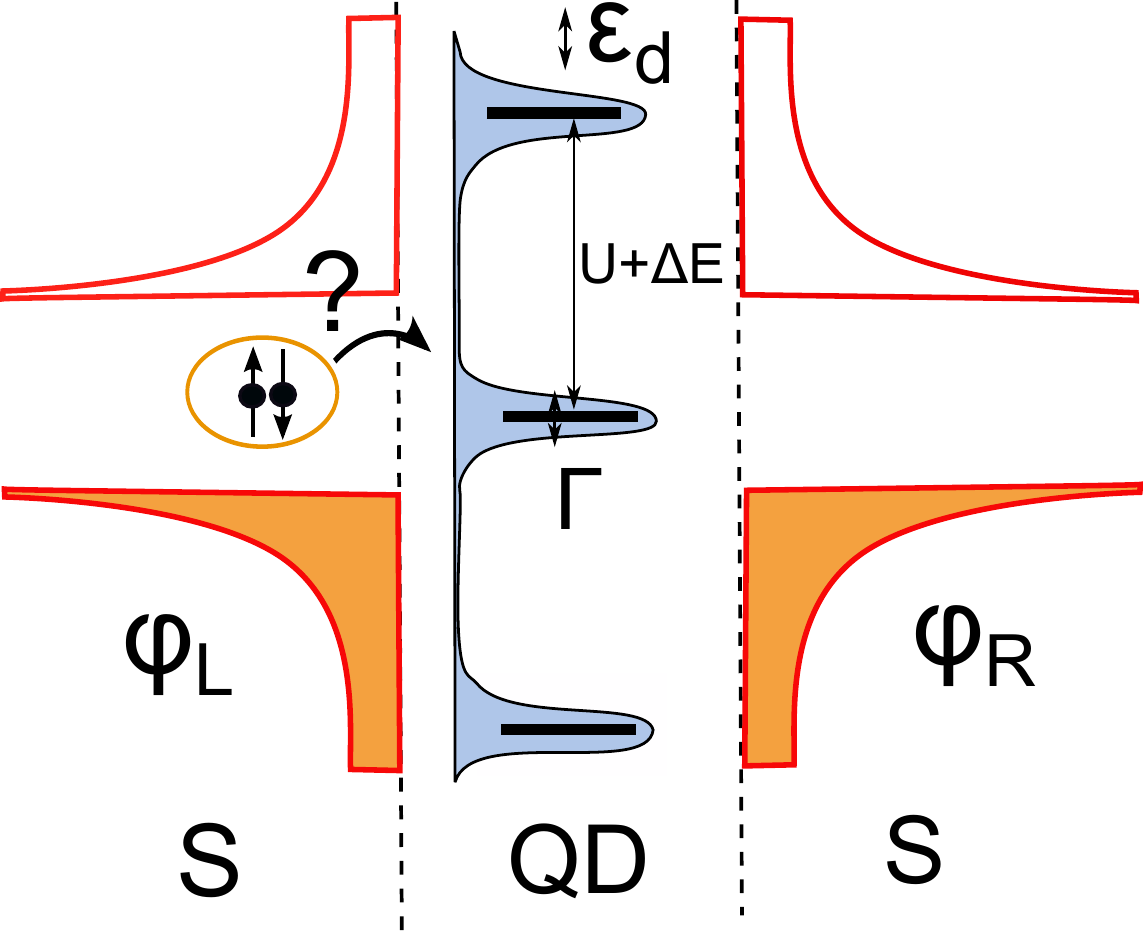}
\end{center}
\caption{Schematic of a quantum dot sandwiched between two superconductors. The superconducting reservoirs are supposed identical except that there is a difference of superconducting phase $\varphi=\varphi_R-\varphi_L$ between them. The energy levels in the quantum dot are represented with their characteristic energy scales: the charging energy U, the energy level spacing $\Delta E$, the width $\Gamma$ of the levels (also the coupling between the dot and the reservoirs) and $\epsilon_d$ the shift of the energy levels, controlled by a gate voltage.}
\label{S-QD-S}
\end{figure}

It has been shown in 1999 that a carbon nanotube can sustain a supercurrent by proximity effect \cite{Kasumov_1999}. A contacted carbon nanotube with barriers at the interface with reservoirs can be seen as a quantum dot, with quantized energy levels and in the Coulomb blockade regime if the temperature is low enough. The energy scales of the system are shown on fig. \ref{S-QD-S}: the charging energy $U$needed to add an electron on the dot, the energy level spacing $\Delta E$ and the width $\Gamma$ of the levels (also the coupling between the dot and the reservoirs). The position of the energy levels can be shifted by $\epsilon_d$, controlled by a gate voltage. 

Concerning the proximity effect, the situation is strongly dependent on the relative values of the parameters of the dot:

If $\Gamma\gg U,\Delta$, in the strong coupling regime, the energy levels in the dot are so broad that they overlap : the charge fluctuates in the dot. 
On the contrary, if $\Gamma\ll U,\Delta$, the coupling is low such that the system is in a strong Coulomb blockade regime: out of the charge degeneracy points, it is not possible to add or remove an electron on the dot. 
 
 The most interesting case is the intermediate regime $\Gamma\approx U\approx\Delta$. In this limit, $U$ is high enough to give rise to Coulomb blockade while $\Gamma$ is sufficiently large to allow the cotunneling processes needed for the development of the Kondo effect. In the normal state, these cotunneling processes allow an electron to enter the dot (and thus to change the electronic occupancy) provided that another one goes out during a time shorter than $\hbar/U$ (the typical tunneling time being $\hbar/\Gamma$). In the superconducting state the situation is more complicated since, instead of a single electron, one needs to make the two electrons of a Cooper pair tunnel through the dot coherently in order to observe a supercurrent. Because of the exclusion principle, the situation will be strongly dependent on the parity of the number of electrons in the dot. In the next section, we give a qualitative image of what happens in this intermediate regime.

\subsection{Parity induced 0-$\pi$ transition}

To transfer a Cooper pair through a quantum dot by cotunneling processes, one needs to break the pair (which is possible during a time $\hbar/\Delta$) and make the electron co-tunnel one by one (during typically the time $\hbar/\Gamma$): this is possible only if $\Delta<\Gamma$. Depending on the parity of the number of electrons in the dot, different processes are possible (see ref. \cite{Spivak1991,defranceschi2010}).

\begin{figure}[htbp]
\begin{center}
\includegraphics[height=6cm]{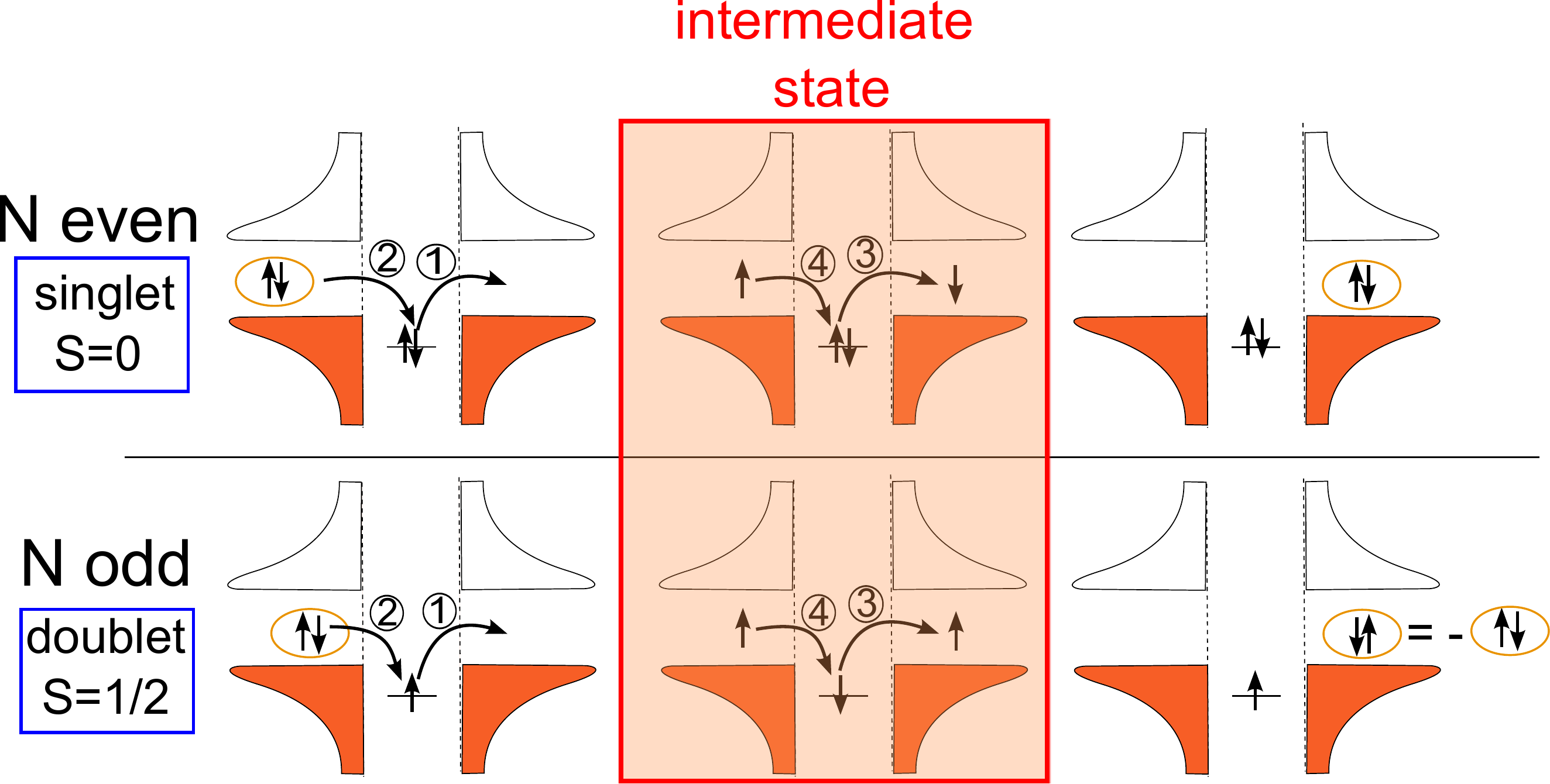}
\end{center}
\caption{Qualitative explanation for the parity induced 0-$\pi$ transition. Depending on the even or odd parity, the co-tunneling processes are different (see text).}
\label{jonction_pi_qual}
\end{figure}

For an even number of electrons, the highest occupied energy levels is filled, as represented on fig. \ref{jonction_pi_qual}, the system is in a singlet state. To transfer a Cooper pair from the left to the right reservoir, one electron from the dot tunnels to the right electrode (1) and is replaced by the electron of the Cooper pair with the same spin (2). In a second step, the electron on the dot with the opposite spin tunnels to the right and is replaced by the second electron of the Cooper pair. At the end, the ordering of the Cooper pair is unchanged.

For an odd number of electrons, the system is in a doublet state. Things are different since there is only one electron, spin up or down, in the highest occupied state. If, in the first step, we inter-exchange electrons of same spin, it will not be possible to reform a zero-spin Cooper pair in the right electrode. We thus have to invert the spins: in the first step a spin down (for example) of the Cooper pair replaces a spin up in the dot. In the second step, spin up of the Cooper pair replaces a spin down in the dot (see fig. \ref{jonction_pi_qual}). At the end, the spin ordering of the Cooper pair formed in the right contact has been inverted.

These spin-flip processes strongly affect the supercurrent, as it has been first pointed out by Kulik in 1966 \cite{Kulik1966}. These processes are of fourth order and, in the doublet state, are the first one leading to a non-zero current. Consequently, the supercurrent is weakened \cite{Novotny2005} and the current-phase relation is dephased by $\pi$ ($I=I_c\sin(\varphi+\pi)=-I_c\sin(\varphi)$): this is a $\pi$-junction \cite{Glazman1989,Spivak1991,defranceschi2010}. This name is given by opposition to the singlet case, where the current-phase relation is a regular one: $I=I_c\sin(\varphi)$.

 The transition from a 0 to a $\pi$ junction is achieved by tuning the dot's occupancy with a gate voltage, yielding a gate-controlled 0-$\pi$ transition. It has been first experimentally observed by Van Dam \textit{et al.} \cite{Vandam2006} in an InAs nanowire QD and Cleuziou \textit{et al.} \cite{Cleuziou2006} in a CNT. In both experiments, the QD is inserted in a superconducting loop so that the phase is controlled by a magnetic field, allowing the measurement of the current-phase relation. Jorgensen \textit{et al.} \cite{Jorgensen2007} measured as well this gate-controlled 0-$\pi$ transition, also in a CNT but without control of the phase.

\subsection{Competition between Kondo effect and superconductivity}

We now focus on Coulomb diamonds corresponding to an odd number of electrons in the presence of the Kondo effect. This effect, well known in dilute alloys \cite{Kondo1964,Anderson1970,Wilson1975}, leads to the screening of the magnetic impurities embedded in a metal by the conduction electrons around the Fermi energy. A quantum dot connected by metallic electrodes and occupied by an odd number of electrons constitutes an artificial realization of Kondo impurity \cite{Pustilnik2004,Cronenwett1998,Nygard2000}. 

 In this system, the Kondo effect manifests as enhanced cotunneling processes which overcome Coulomb blockade. This gives rise to a resonance in the density of states at the Fermi level, $i.e.$ as a conductance resonance at zero bias, as well as a screening of the local magnetic moment of the dot. In other words, the Kondo effect transforms the doublet state induced by Coulomb blockade in a Kondo singlet state. The characteristic energy associated to this effect defines the Kondo temperature, and can be expressed as \cite{Haldane1978}:
 \begin{equation} 
 \label{Tk}
 k_BT_K=\sqrt{\frac{U\Gamma}{2}}e^{\frac{\pi}{2}\epsilon_d(\epsilon_d+U)/(U\Gamma)}
 \end{equation}

When the electrodes of a Kondo quantum dot turn superconducting, different regimes can take place \cite{defranceschi2010}:
\begin{itemize}
\item If $\Delta\gg T_K$, there is no electron around the Fermi level in a range of energy $T_K$ that is able to participate to the Kondo screening. The situation is not modified compared to the case described above: this is a $\pi$-junction.
\item In the $\Delta\ll T_K$ case, the Kondo correlations coexist in the system with superconducting ones. The cotunneling processes are enhanced, making the transfer of Cooper pairs easier, and favoring the formation of a Kondo/BCS singlet state. Both Kondo effect and superconductivity cooperate so that a 0-junction is recovered, with possibly a very high supercurrent.
\item  Intermediate regime $\Delta\approx T_K$ : how does the system transit from 0 to $\pi$-junction in this regime of strongest competition? We will see that in this regime, the superconducting phase plays a role in the transition, as predicted first by Rozhkov \textit{et al.} \cite{Rozhkov2001} in 1999 and Clerk \textit{et al.} in 2000 \cite{Clerk2000}.
\end{itemize}  

This 0-$\pi$ transition driven by the ratio $\Delta/T_K$ was predicted some time ago by Glazman and Matveev, in 1989 \cite{Glazman1989}. The first occurrence of this interplay came in 2002 by Buitelaar \textit{et al.} \cite{Buitelaar2002}, who proved that the conductance at zero bias in the superconducting state is suppressed if $\Delta>T_K$ and enhanced otherwise. Then, in 2009, this result has been supported by critical current measurements in a carbon nanotube by Eichler \textit{et al.} \cite{Eichler2009} (only the critical current was measured, not the current-phase relation).

Note that the problem has also be tacked through the spectroscopy of the Andreev bound states, as for example in ref. \cite{Pillet2010,Pillet2013,Kim2013,Chang2013,Lee2014}. These experiments enable in particular to visualize the crossing of the Andreev levels as a function of the different parameters.

\subsection{Phase dependence of the ABS, current-phase relation and phase-driven 0-$\pi$ transition.}

\begin{figure}
\begin{center}
\includegraphics[height=7cm]{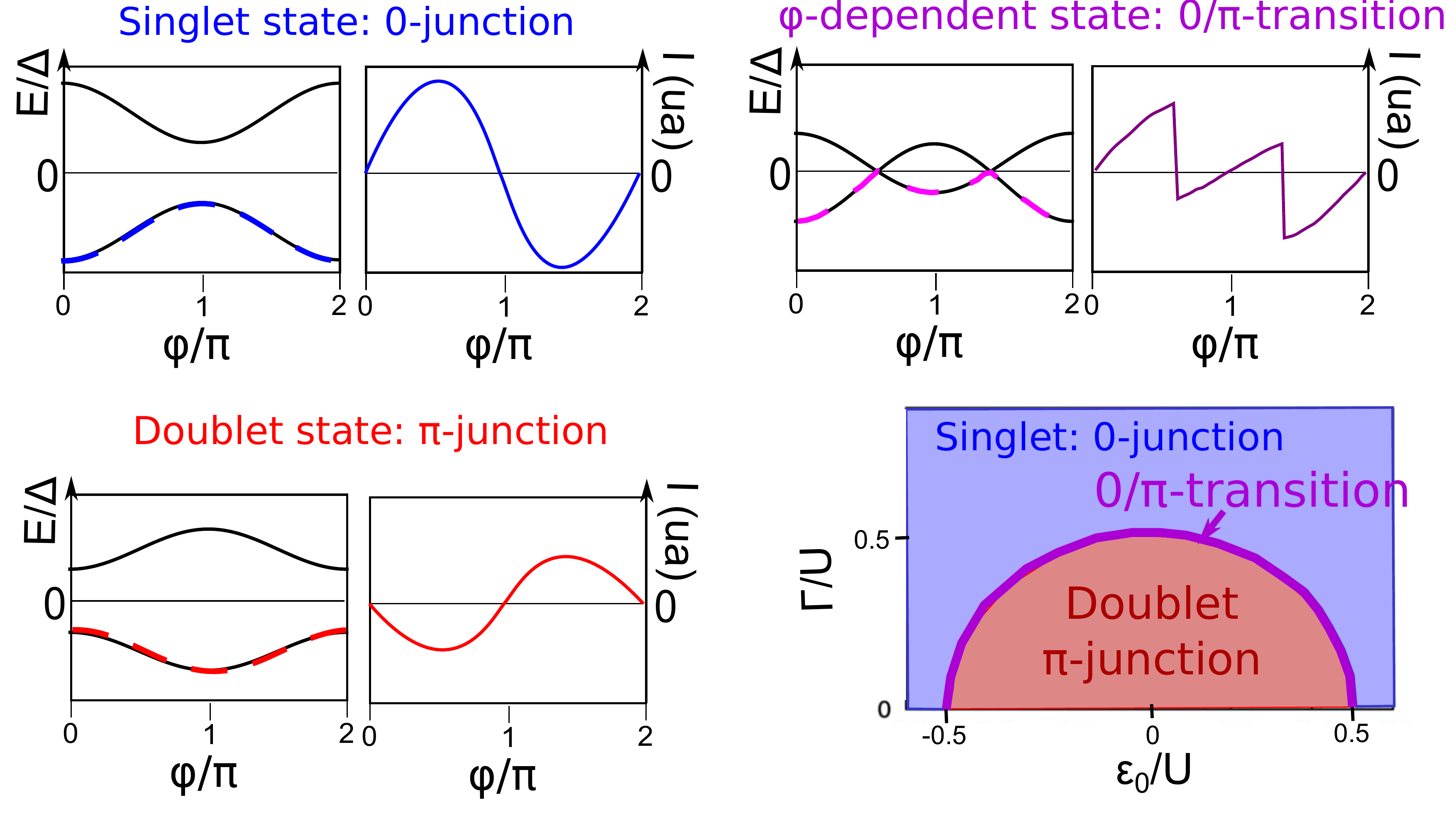}
\end{center}
\caption{Andreev Bound States and corresponding current-phase relation in the three regions of the phase diagram (in the bottom right): the singlet state (in blue), doublet state (red), and at the transition (purple line).}
\label{phi-dep_ABS_CPR}
\end{figure}

To understand the current-phase relation at the 0-$\pi$ transition, we have to understand the phase dependence of the ABS in the system. They can be calculated in the general case taking into account the Kondo correlations, as for example in ref. \cite{Meng2009}. On fig. \ref{phi-dep_ABS_CPR} are represented qualitatively the phase-dependence of the Andreev Bound States , inspired by ref. \cite{Vecino2003} in three different regimes, as well as the derivative as a function of the phase of the ones below the Fermi energy, giving the dominant contribution to the supercurrent at zero temperature.
\begin{itemize}
\item The singlet state corresponds to the blue part on the phase diagram, where $T_K>\Delta$ such that the Kondo effect cooperates with the superconductivity to form the singlet-state. The most striking difference with a standard SNS junction is that the degeneracy of the ABS is broken : each ABS is split in two because of the Coulomb interaction \cite{Vecino2003}. Moreover the ABS are detached from the continuous spectrum above the gap $\Delta$. However, the sign of the current is the same as in a standard junction: this is a 0-junction.
\item The doublet state, where $T_K<\Delta$ such that the Kondo effect does not survive to superconductivity, corresponds to the red region. Then the bound states have been exchanged: the sign of the supercurrent is reversed, this is a $\pi$-junction.
\item The third case is the most intriguing, corresponding to the purple line on the phase diagram, where the Kondo effect and superconductivity are of the same order of magnitude ($T_K\approx\Delta$). There, the inner ABS's cross at the Fermi energy for some values of superconducting phase (see fig. \ref{phi-dep_ABS_CPR}) such that the ground state around $\varphi=\pi$ is a doublet and a singlet around $\varphi=0$. The consequence on the CPR is spectacular: it looks like a $\pi$-junction around $\varphi=\pi$ and to a 0-junction around $\varphi=0$. Since the system transits from a doublet to a Kondo singlet varying the superconducting phase, we can see the situation as a phase-dependent Kondo screening.
\end{itemize} 

This doublet to singlet transition driven by the phase can also be shown using renormalization group theories \cite{Karrasch2008,choi2004} and Quantum Monte Carlo \cite{Siano2004,Luitz2010}. 

However, before our work, this kind of current-phase relation had not been measured. The CPRs measured by Maurand \textit{et al.} \cite{Maurand2012} at the 0-$\pi$ transition exhibit some anharmonicities which may be related to this phenomenon, but the symmetry of these curves is problematic: they are not odd functions of the superconducting phase as they should be in absence of any breaking of the time reversal symmetry. 

The aim of this work is to measure accurately and systematically the current-phase relation at this 0-$\pi$ transition in order to demonstrate the driving of the 0-$\pi$ transition by the superconducting phase.

\subsection{Josephson effect in a two-level quantum dot}
\label{Josephson_2L}	
	
Until now, we have considered a single-level (SL) quantum dot, where the measurement of the CPR (0 or $\pi$-junction) is related to the state of the system (singlet or doublet). But, in clean carbon nanotubes, each energy level is nearly orbitally degenerated in addition to the spin degeneracy, such that they generally form two-level (2L) QD. This may affect dramatically the Josephson effect.

\begin{figure}[htbp]
\begin{center}
\includegraphics[height=3.5cm]{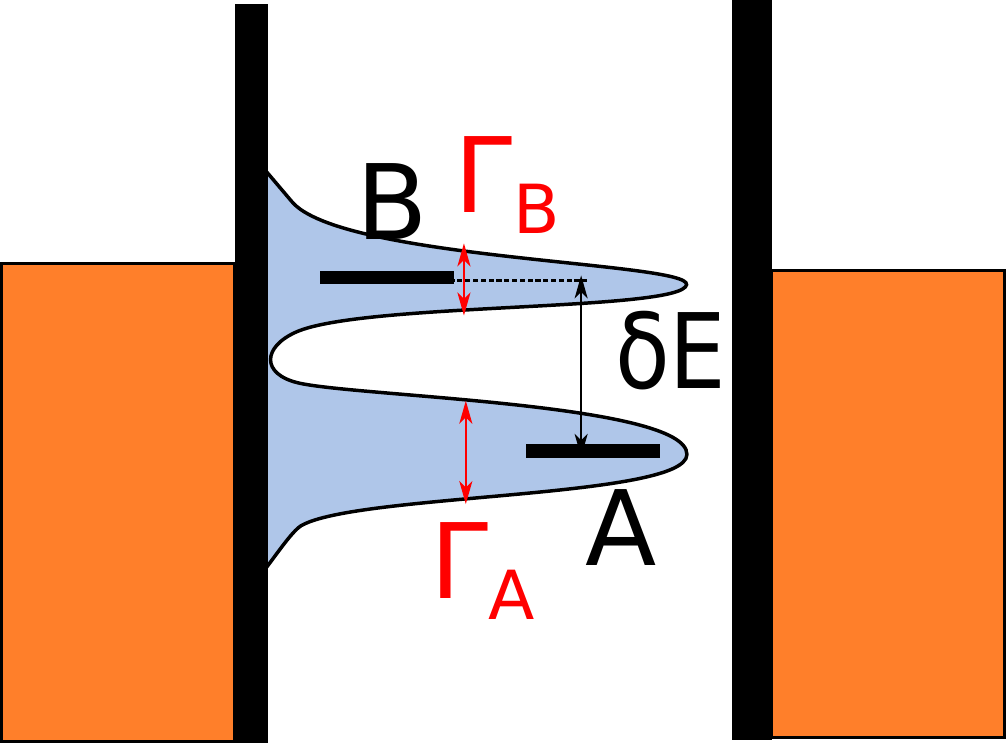}
\hspace{1cm}  
\includegraphics[height=3.5cm]{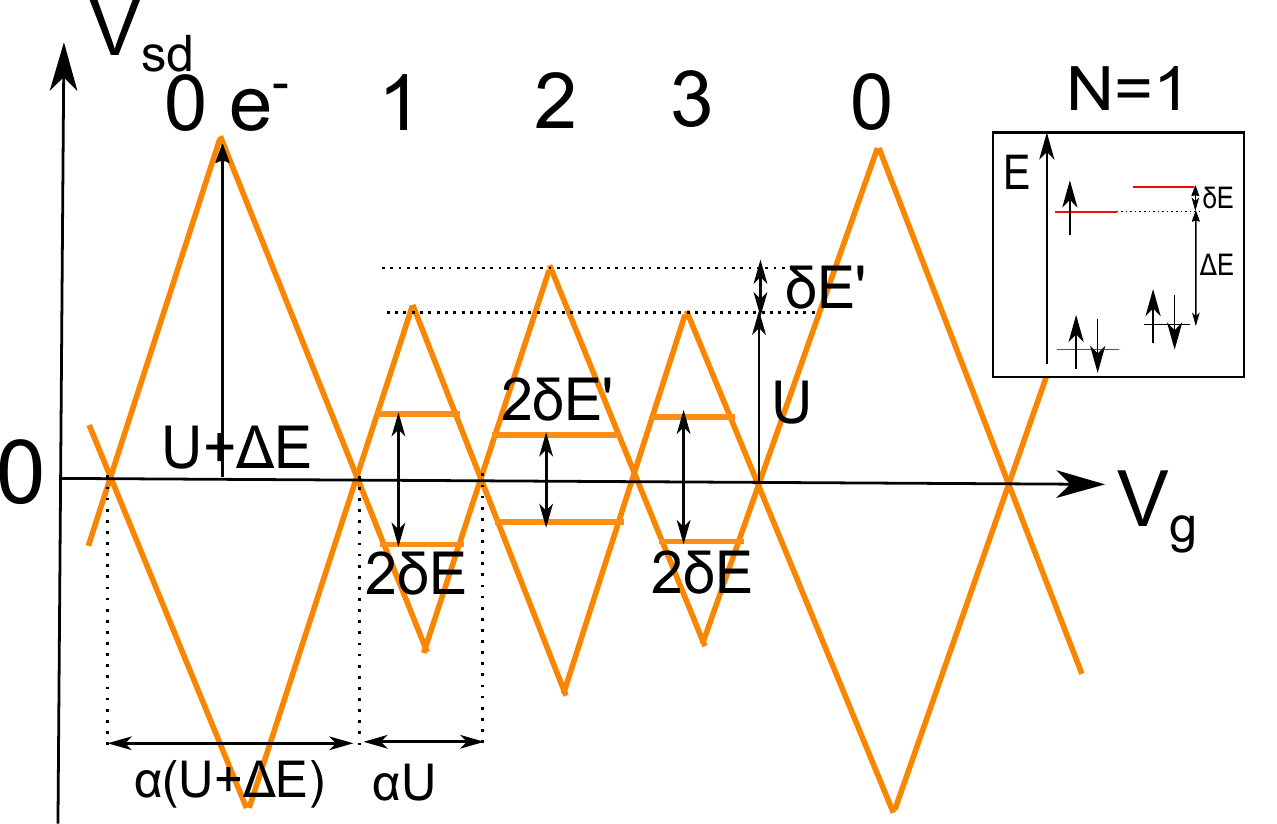}
\end{center}
\caption{Left : Schematic of a two-level quantum dot such as the one we are interested in. The two energy levels, called A and B, are separated by an energy $\delta E$. They may be differently coupled to the reservoirs. Right :  Typical stability diagram expected in a clean CNT, of charging energy $U$, where the energy
levels are separated by $\Delta E$ and the orbital degeneracy is lifted by $\delta E$ . The occupancy is indicated on the top of each Coulomb diamond. We call $\alpha$ the proportionality factor between the energy level and the applied gate voltage Vg . The spacing between the inelastic cotunneling peaks in the double
occupancy (N=2) is called $2 \delta E'$. $\delta E' \neq \delta E$ because of exchange interactions \cite{babic2004}. Inset: corresponding electronic configuration for N=1.}
\label{2L-QD_schema}
\end{figure}

 The situation is summarized on fig. \ref{2L-QD_schema}. We want to induce a supercurrent in a QD where the gate voltage is chosen such that two levels may participate to the transport of Cooper pairs: these two levels are called A and B and are separated by an energy $\delta E$ (which represents the breaking of the orbital degeneracy). They may be differently coupled to the reservoirs and we call respectively $\Gamma_A$ and $\Gamma_B$ these couplings (they may be different for left and right reservoirs, but we will neglect this point in the following of this part).

As stated above, the $\pi$-junction in a SL-QD with an odd number of electrons is possible since the Cooper pair cannot be transfered without spin-flip during the cotunneling. But if other levels (empty or full) are available, this is not true anymore  since a new path become available for electrons. This has been experimentally pointed out in 2006 by Van Dam \textit{et al.} \cite{Vandam2006}. In this experiment, the Josephson effect has been measured in an InAs nanowire where the Kondo effect was negligible and several levels participated to transport. They measured the supercurrent at a fixed phase $\varphi=\pi/2$ as a function of the gate voltage (this quantity can roughly be seen as the critical current for harmonic current-phase relations, with a positive sign for 0-junctions and negative for $\pi$-junctions). The supercurrent in some diamonds is quite surprising: it becomes negative for an even number of electrons. In addition to that, the supercurrent is not symmetric compared to the center of the diamond. Such a behavior was actually predicted in 1998 by Shimizu \textit{et al.} \cite{Shimizu1998a} and is due to the participation of several levels to transport. In this work the authors have shown that in a multi-level regime, using a random distribution of energy levels with different couplings, both 0 and $\pi$-junctions are possible for both odd and even occupancies. 

To our knowledge, there is no other measurement of Josephson effect in a multi-level QD (except in ref. \cite{Szombati2016}, but this question is not raised explicitly). It seems that all the previous measurements of supercurrent realized in CNT were in the SL regime ($\delta E\gg U,\Gamma$). In the next section, some theoretical predictions in the two-level regime are presented.

\subsubsection{Two-level regime predictions without Kondo effect}

The supercurrent in a two-level quantum dot has been calculated by Shimizu \textit{et al.} using an Anderson-like Hamiltonian and neglecting the Kondo correlations\cite{Shimizu1998a}. For each co-tunneling event, there are \textit{a priori} $24=4!$ sequences possible to transfer a Cooper pair through the QD.  The sign of the contribution to the current of each sequence is given by the number of electron permutations (\textit{i.e.} the number of processes with spin-flip) and the sign of the wave-function (see ref. \cite{Shimizu1998a} and supp. informations of ref. \cite{Vandam2006}).

In the single level regime, when there is one electron in the dot, only six of them are available, and they contribute negatively to the current. When there are two electrons in the dot, only some processes giving positive contributions to the currents are allowed. In a two-level QD, a second path is available, such that the twenty-four processes are \textit{a priori} possible for both even and odd occupancies. That is why a $\pi$-junction is not necessary related to a doublet state.
 
Still without considering the Kondo effect, Yu \textit{et al.} focused on the carbon nanotube case, with two spin-degenerate levels \cite{Yu2010a}. Two parameters, specific to CNT, are taken into account: the small energy $\delta E$ between the two levels, and a new parameter, $T_2$, which quantifies how much the orbital degree of freedom is conserved during tunneling: $T_2=0$ if it is conserved and is non-zero otherwise. 
The main influence of $\delta E/\Delta$ is to increase the size of the N=2 diamond. The influence of orbital mixing ($T_2\neq 0$) is more interesting: it suppresses the $\pi$ behavior at $N=2$.

Note that Droste \textit{et al.} \cite{Droste2012} and Karrasch \textit{et al.} \cite{Karrasch2011} have also calculated the supercurrent in double dots in series, a system \textit{a priori} different from a CNT where both levels are connected to both reservoirs. The results are thus not directly transposable.

Note as well that, in all these articles, no current-phase relations are presented in the ML-regime. Current-phase relations are however presented in the article by Lee \textit{et al.}, but it concerns only the 2L-QD occupied by two electrons.

\subsubsection{What about the Kondo effect?}
When the Kondo correlations are taken into account, the situation becomes much more complex. It has for example been taken into account in an article by Lee \textit{et al.} \cite{Lee2010}, but only in the case of a double occupancy of the 2L-QD.

Zazunov \textit{et al.} \cite{Zazunov2010} have tackled more specifically the case of the SU(4) Kondo effect (for N=1 occupancy), Lim \textit{et al.}\cite{Lim2011} did it in presence of spin-orbit coupling. The calculations provide also a rich phase diagram. 

\subsection{0-$\pi$ transitions}
\label{sec_0pitransition}
In the single-level regime, we have presented a transition from a 0 to a $\pi$-junction, originating from a crossing of Andreev levels, induced by an interplay between Coulomb interactions and the superconducting correlations, helped by the Kondo effect. However 0-$\pi$ transitions can happen in other situations that are reviewed here.

\subsubsection{Back to the $\pi$-junctions}
In order to generalize the notion of $\pi$-junction, we can consider the free energy F of the junction, from which the Josephson current is calculated with  $I=-\frac{2e}{\hbar}\frac{\partial F}{\partial \varphi}$ \cite{Golubov2004}.
%\begin{eqnarray}
%E_J=\int\limits_{0}^{t}I(t')V(t')dt=\int\limits_{0}^{t}\frac{\hbar %I_c}{2e}\cos(\varphi(t'))\frac{d\varphi}{dt'}dt\\
%E_J=\frac{I_c\hbar}{2e}\left(1-\cos(\varphi)\right)
%\end{eqnarray}	

\begin{figure}
\begin{center}
\includegraphics[height=5cm]{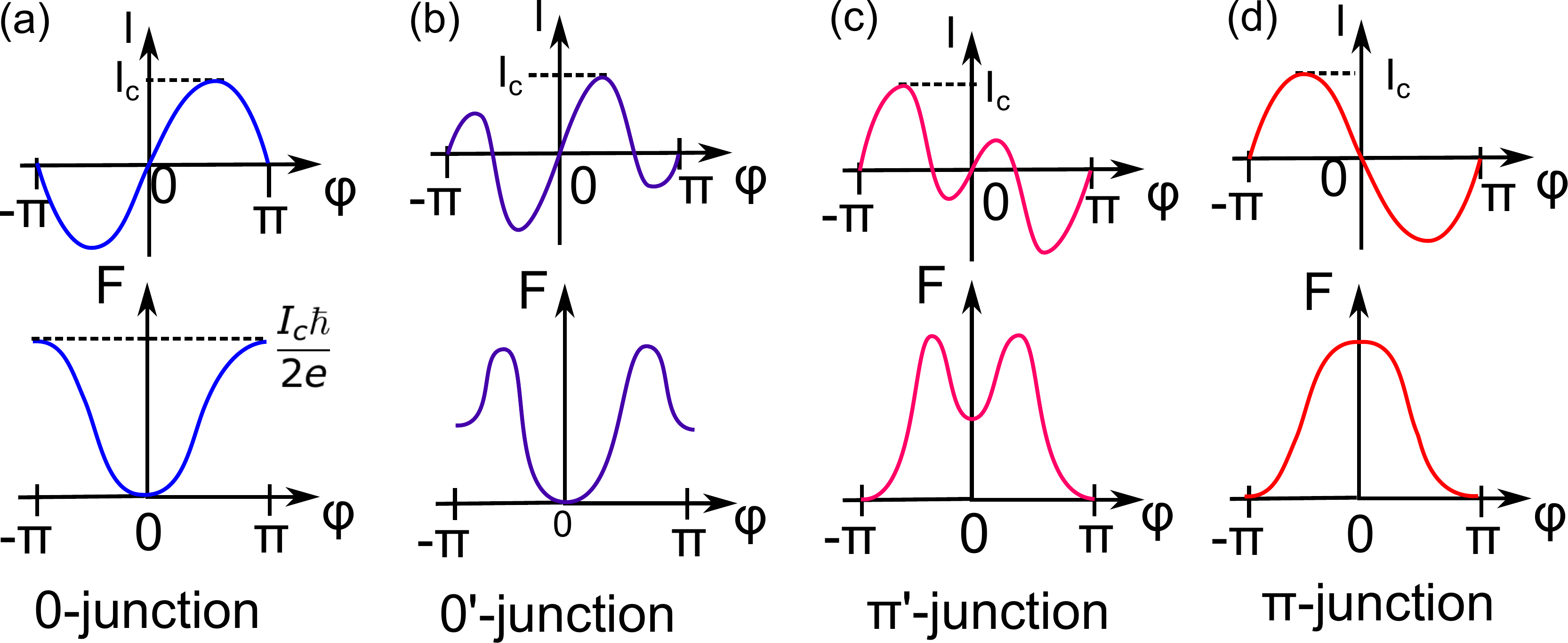}
\end{center}
\caption{Current-phase relation and free energy F (see text) for four different kinds of Josephson junctions.}
\label{EJosephson}
\end{figure}

If the current-phase relation is $I=I_c\sin(\varphi)$ (+ harmonics), the minimum of $F$ is at $\varphi=0$: this is why we call it a 0-junction (see fig. \ref{EJosephson} (a)). On the other hand, for a CPR $I=I_c\sin(\varphi+\pi)$ (+ harmonics) the minimum energy is at $\varphi=\pi$: this is a $\pi$- junction (fig. \ref{EJosephson} (d)) \cite{Golubov2004}. What about the CPRs predicted at the single-level 0-$\pi$ transition, as represented on fig. \ref{phi-dep_ABS_CPR}? We argued that the CPR has a 0 or a $\pi$ behavior depending on the value of the superconducting phase, and is thus neither 0 nor $\pi$ state. The situation is depicted on fig. \ref{EJosephson} (b) and (c), taking into account a temperature broadening : depending on the proportion of 0 and $\pi$ behaviors, the global minimum of the free energy is at $\varphi=0$ or $\pi$. But there is as well a local minimum at respectively $\varphi=\pi$ and $0$: that is why they are called a $0'$ or $\pi'$ junctions \cite{Siano2004}.
In the SL 0-$\pi$ transition described above, the system thus transits from 0 to $\pi$ through 0' and $\pi'$ states. Is it general to all 0-$\pi$ transitions? Does it depend on their physical origin? What are the different kinds of 0-$\pi$ transitions? That's the questions we try to address now.
	
\subsubsection{Various 0-$\pi$ transitions}

\paragraph{Single level Quantum Dot 0-$\pi$ transition}
As stated above, this 0-$\pi$ transition is the consequence of a transition from a doublet to a Kondo/BCS singlet, driven by the ratio $\Delta/T_K$. In this context, it is worth noting that the critical current is expected to be non-zero at the transition, since the system transits through 0' and $\pi'$ states. 

\paragraph{Two-level Quantum Dot 0-$\pi$ transition}
In a two-level QD, transitions from $\pi$ to 0-junctions may happen when the contributions of the various processes (those leading to positive and negative current) vary with the gate voltage. How is the CPR at the transition? Lee \textit{et al.} answer this question in the case of a double occupancy of the 2L QD. When the 0/$\pi$ transition corresponds to a transition between two different ground states (singlet/triplet transitions), it goes through intermediate $0'$ and $\pi'$ states (with the corresponding anharmonic CPR). However, when the 0-$\pi$ transition does not involve any change of magnetic state of the dot, the amplitude of the supercurrent simply goes to zero. To our knowledge, there is no generalization of this work for single occupancy of the 2L QD.

\paragraph{SFS $\pi$-junctions/Zeeman $\pi$-junctions}
The most famous way of fabricating a $\pi$-junction is to make a ferromagnetic Josephson junction, a thin ferromagnetic layer sandwiched between two superconductors (SFS). In this context, this is equivalent to applying a Zeeman magnetic field.
In a ferromagnetic material or in presence of a Zeeman field, it exists an exchange energy $E_{ex}$ between the spins up and down (see refs. \cite{Buzdin2005,Eschrig2011} for SFS junctions or \cite{Yokoyama2014} for Zeeman field $\pi$-junctions). This exchange energy induces a phase shift between the electron and the hole of the Andreev pair after propagation in the junction of length $d_f$: $\Delta \phi=2E_{ex}d_f/(\hbar v_F)$. It follows an oscillation of the superconducting order parameter, leading to a $\pi$-junction when $\Delta \phi \in [\pi/2,3\pi/2]$ \cite{Kontos2002}.
                       
 Another way to see the phenomenon is to remark that the spin-degeneracy of the ABS is lifted by the exchange/Zeeman field \cite{Yokoyama2014}. Depending on its amplitude, the position of the ABS is modified, potentially leading to $\pi$-junctions or  0'/$\pi'$-junctions in the case of ABS crossings. 
 
 To the best of our knowledge, there exists only very few measurement of Zeeman $\pi$-junctions, observed in a bismuth nanowire, benefiting of a large g-factor \cite{Murani2017} or HgTe quantum wells \cite{hart_controlled_2017}. However, in SFS junctions, the phenomenon has been largely investigated and is still an intense subject of research, in relation with superconducting spintronics. The 0-$\pi$ transition in SFS junctions, driven by the thickness of the F layer, has been observed by Kontos \textit{et al.} in 2002 \cite{Kontos2002} through critical current measurements, followed in 2003 by the phase dependence \cite{Guichard2003}. If the temperature T is of the order of the exchange energy, the 0-$\pi$ transition can also be driven by the temperature \cite{Ryazanov2001}. The current-phase relation has been measured at this T-driven 0-$\pi$ transition by Frolov \textit{et al}. \cite{Frolov2004}, giving an interesting result: at the transition, the amplitude of the CPR ($i.e$ the critical current) vanishes, or at least becomes too small to be measured. This result is in conflict with most theories predicting that, even though the first harmonic vanishes at the transition, it should remain a least a double harmonic contribution, so that the critical current is non-zero at the transition \cite{Samokhvalov2015}. This second harmonic has been indeed detected by Shapiro steps measurements done in 2004 \cite{Sellier2004}. If we consider the transition as originating from a crossing of ABS, there should obviously be higher harmonics in the CPR at the transition, but they may be difficult to measure at the temperatures at which the transition happens (of the order of 3K).

\paragraph{Controllable Josephson junctions} 
If an SNS junction is thermally excited, the ABS above the Fermi energy can be populated. Then, they  participate as well to the supercurrent giving a negative contribution: the amplitude of the supercurrent can be controlled \cite{Morpurgo1998} or even reversed \cite{Baselmans1999}. Because of the phase-dependence of the ABS, it is easier to populate the excited levels for superconducting phases around $\varphi=\pi$. That is why 0' and/or $\pi'$ phase are predicted and measured \cite{Baselmans_PRL2002}. It is interesting to note that the measured current-phase relations at the transition have the same unexpected even symmetry as the one measured by Maurand \textit{et al.}.

\subsubsection{Quantum phase transitions?}
\label{sec_QPT}	
We have presented a number of 0-$\pi$ transitions, whose common characteristic is a sign reversal of the supercurrent. These sign reversals have a wide variety of origins: singlet-doublet transition, thermal excitation or a Zeeman/exchange splitting of the ABS, two-level regime in QDs. 

Is it possible to classify these transitions among the (quantum) phase transitions? 

A quantum phase transition (QPT) is a phase transition happening at zero temperature, induced by the variations of a parameter different from temperature \cite{Vojta2006}. As any phase transition, it results from the competition between different ground states of the system, leading to different macroscopic phases. In this respect, the 0-$\pi$ transition induced by thermal excitation by Baselmans \textit{et al.} \cite{Baselmans_PRL2002} is not a QPT.

Just like classical phase transitions, QPTs can be classified in first order and second (or higher) order transitions. First order ones result from a level crossing, such that the two states coexist during the transition. In a second order QPT, the transition is continuous and driven by quantum fluctuations instead of thermal ones \cite{Rau2013}: there exists a quantum critical region which is described by critical exponents, providing a very rich physics. 

Speaking of quantum phase transition in a quantum dot is questionable. Is the reversing of the sign of the supercurrent a change of phase? However, if it is the manifestation of a transition from a doublet to a singlet ground state, then this terminology is clearly suitable provided that this ground state is a collective one. 

Some quantum phase transitions have been investigated in quantum dots. Mebrahtu \textit{et al.} \cite{Mebrahtu2012} studied resonant tunneling in a Luttinger liquid formed in a CNT between two barriers, where a second order QPT occurs varying the symmetry of these barriers. Roch \textit{et al.} \cite{Roch2008} have measured an infinite order QPT (Kosterlitz-Thouless transition) in a single molecule QD, where the system transits from a singlet to a triplet ground state varying the gate voltage. In both cases (singlet and triplet), the Kondo effect is involved, inducing a correlated state between the dot and the leads: it is appropriate to speak of QPT. Another example of QPT of second order is given by the investigation of the crossover between one and two-channel Kondo effect by Iftikhar \textit{et al.} \cite{Iftikhar2015}.

\begin{figure}[htbp]
\begin{center}
\includegraphics[height=7cm]{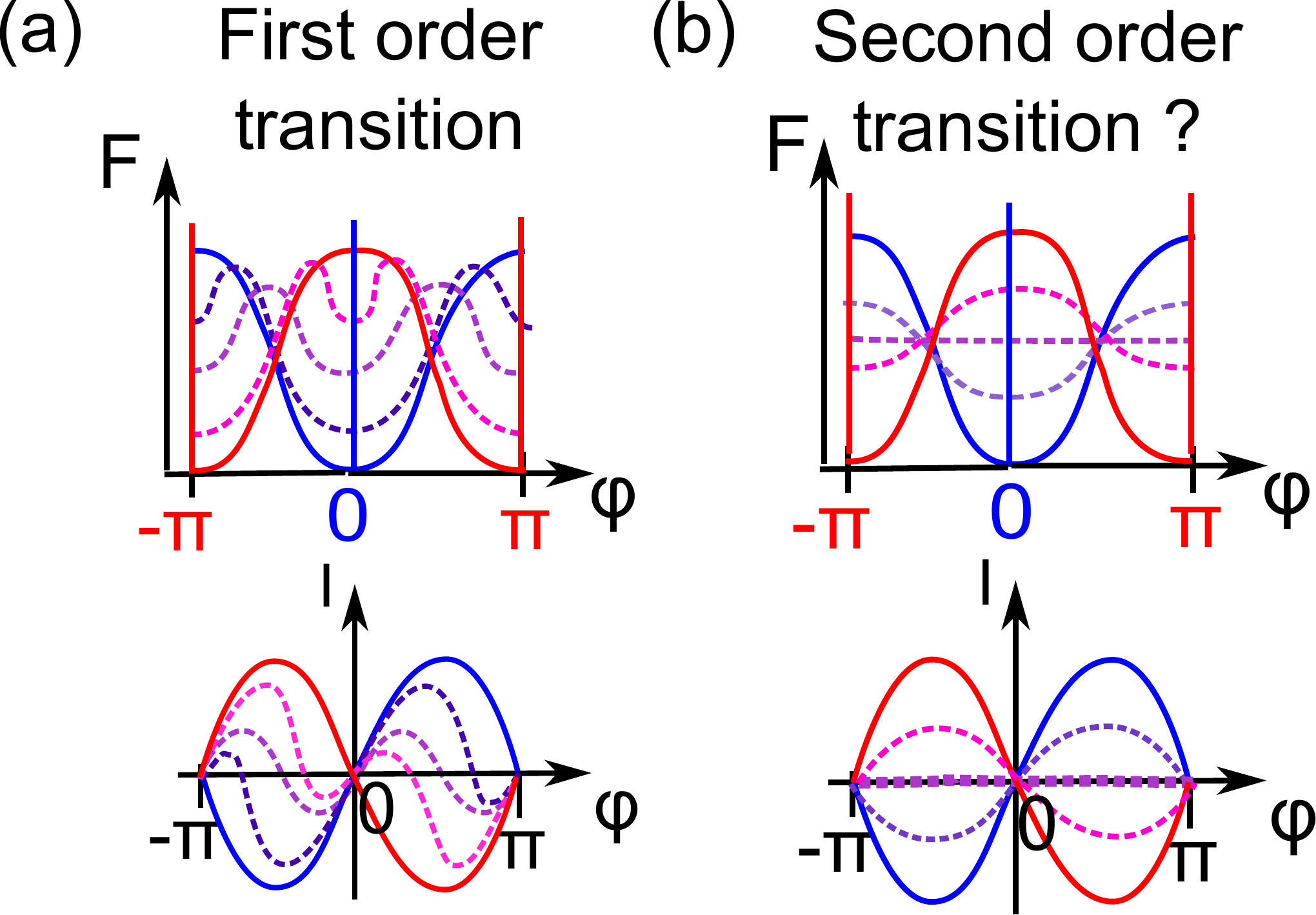}
\end{center}
\caption{Qualitative representation of free energy and the current-phase relation expected for two kinds of 0-$\pi$ transitions: (a) first order transition, as predicted for the 0-$\pi$ transitions due to level-crossing and (b) an hypothetic second order transition. In the first order case, at the transition, both 0 and $\pi$ states coexist at different $\varphi$, the critical current never vanishes and the CPR is anharmonic. In the second order case, the amplitude of the CPR decreases until 0, where the sign change occurs, and increases again in the other state.}
\label{1-2_order_transition}
\end{figure}

In the case of the 0/$\pi$ transition in a single-level QD, the use of the term quantum phase transition is reasonable. It is clearly due to a crossing of Andreev levels, and is thus a first-order QPT \cite{Maurand2012}. Typical free energies F and the corresponding CPRs are represented on fig. \ref{1-2_order_transition} (a): at the transition, F has two minima, one at 0 and the other at $\pi$ (one of these minima can be local and the other global). That is why the CPR is strongly non-harmonic and its critical current never vanishes. The transition is discontinuous since, between the 0 and $\pi$ states, one can find mix states (0' and $\pi'$). Note that there is no quantum criticality associated to this transition.

0-$\pi$ transitions in SFS junctions and induced by a Zeeman field are also due to a level-crossing, the CPR at the transition should be similar to fig. \ref{1-2_order_transition} (a). 

The situation is more complex in two-level quantum dots. According to Lee \textit{et al.} who calculated the case of double occupation, while 0-$\pi$ transitions are the consequence of a transition from a magnetic state to the other (singlet, doublet, triplet), the system makes a transition through the intermediate states 0' and $\pi'$. But as soon as the 0-$\pi$ transition does not involve a change of ground state of the system itself, there is no intermediate CPR and the critical current vanishes at the transition, as represented on fig. \ref{1-2_order_transition} (b). 
Then the amplitude of the CPR decreases until vanishing at the transition, and increases again but with a sign reversal. In this case, there is no intermediate state 0' or $\pi'$, it is the kind of transition that would be expected for a second order phase transition. Does it mean that these transitions without change of magnetic state are second order transitions or simply that they are not anymore quantum phase transitions? Is it possible to extend this result to single occupancies? According to us, these questions remain open.

\section{Materials and Methods}
\subsection{Principle of the measurement}

\begin{figure}
    \begin{center}
    \includegraphics[height=5cm]{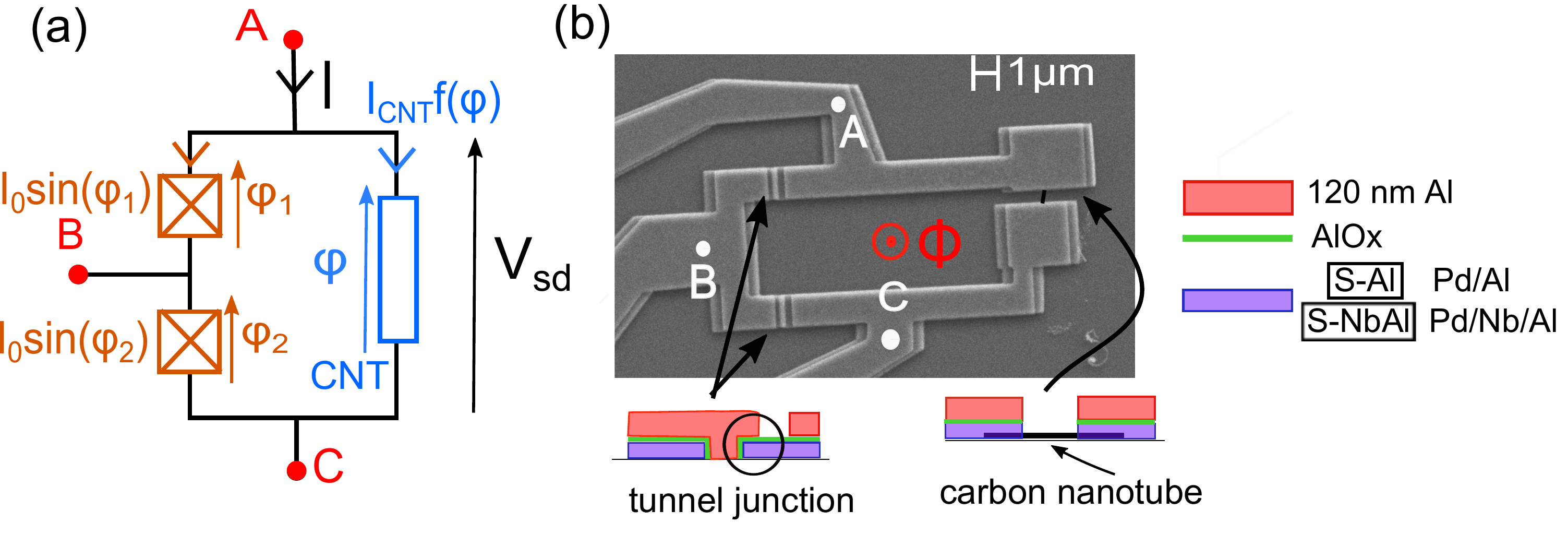}
    \end{center}
    \caption{(a) Schematic of the measured asymmetric SQUID, containing two reference JJs in parallel with a CNT based Josephson junction. To phase bias the CNT junction, a magnetic flux $\Phi$ is applied with a magnetic field perpendicular to the SQUID. (b) Scanning electron microscopy image of the sample, with the layers constituting the tunnel junctions and the contacts of the CNT. Depending on the samples the first layer is made of Pd(7 nm)/Al(70 nm) (sample S-Al) or Pd(7~nm)/Nb(20~nm)/Al(40~nm) (sample S-NbAl).}
	\label{sample}
\end{figure}

We present now our measurements of current-phase relation in a carbon nanotube quantum dot. The principle of the measurement, as well as the specific experimental setup used in this work is described in refs. \cite{Delagrange2015,Delagrange2016,DelagrangePhD}. The CNT QD Josephson junction is embedded in an asymmetric SQUID (Fig.\ref{sample}). We extract the current-phase relation of the CNT QD junction by measuring the switching current of the SQUID as a function of magnetic flux.

\subsubsection{Sample fabrication}

The CNT are first grown by chemical vapor deposition on a silicon wafer covered by an oxide layer. We chose to use a doped silicon wafer so that it can play the role of a back-gate for the nanotube. 
	    
The nanotube contacts and the tunnel junctions are made during one unique lithography step. A first aluminum-based multi-layers is deposited with an angle of 15°, is oxidized under oxygen and is covered by a second layer of aluminum (120 nm) (fig. \ref{sample}). The contacts of the nanotube are separated by a distance $L=400\mathrm{~nm}$. In this work we have measured samples with contacts made of a niobium layer between palladium and aluminum ($\Delta_{PdNbAl} = 170 \pm 5 {~µeV}$, called S-NbAl) : Pd(7~nm)/Nb(20~nm)/Al(40~nm). The sample measured are similar to the one represented on fig. \ref{sample} (b).

The sample is then cooled down in a dilution refrigerator of base temperature 50 mK. The phase difference across the CNT-junction $\varphi$ is controlled applying a magnetic field $B$ perpendicularly to the sample. The magnetic flux enclosed by the loop is $\Phi=B\times S$, with $S\approx 40\mathrm{~\mu m}^2$ the loop area.

\subsection{Characterization of the sample in the normal state}

The samples are first characterized in the normal state.  To do so, we measure the differential conductance $dI/dV_{sd}$ of the sample as a function of the bias voltage $V_{sd}$ and the gate voltage $V_g$ in the normal state. This is done using a standard lock-in-amplifier technique and applying a magnetic field large enough to suppress the superconductivity in the contacts. We explore different Coulomb diamonds, at different gate voltages,  which correspond to different filling factor and parameters (Table \ref{table1}, \cite{Delagrange2015,Delagrange2016}).

\section{Current-phase relation measurement in a CNT quantum dot}

We now restore superconductivity, by switching off the magnetic field, so that we can measure the current-phase relation (CPR) of the QD JJ. In the following we focus on a sample with S-NbAl contact. A more detailed study can be found in ref. \cite{Delagrange2015,Delagrange2016}. 

\subsection{Current-phase relation in the single-level regime}

In this part, we discuss the phase dependence of the supercurrent in the single-level regime, focusing on one sample with occupation number $N=1$ (called I to be consistent with \cite{Delagrange2016}).
	
\begin{figure}
	  \begin{center}
	   \includegraphics[width=16cm]{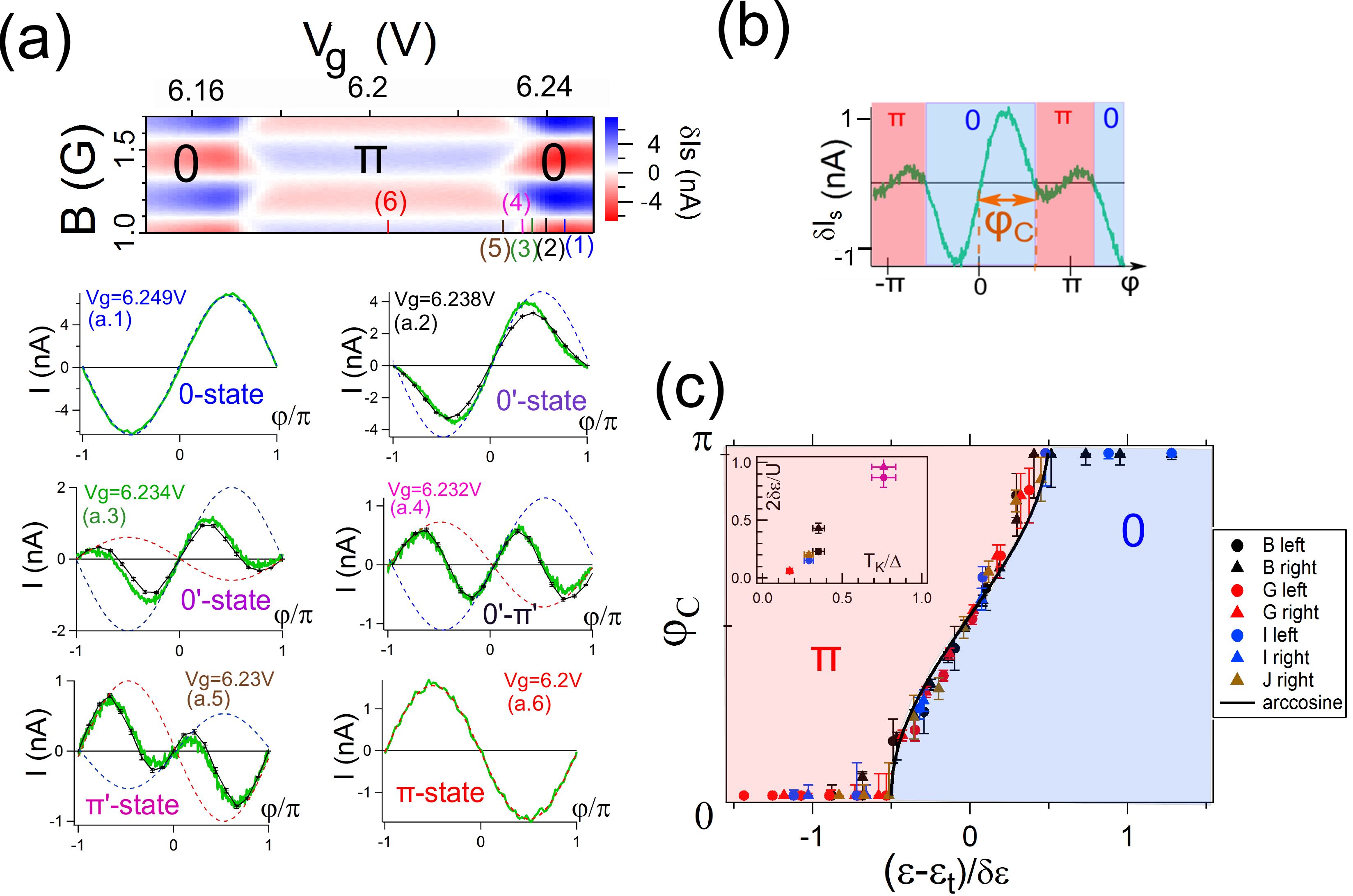}
	   \end{center}
	   \caption{(a) Modulation of the switching current of the SQUID $\delta I_s$, proportional to the CPR,  as a function of the magnetic field B and the gate voltage $V_g$. Vertical cuts at the 0-$\pi$ transition are represented for ridge I, showing the whole transition. The dashed lines are guides to the eyes and represent the contributions of the singlet (0-junction, in blue) and the doublet state ($\pi$-junction, in red). (b)  Definition of the critical phase $\varphi_C$ such that the CPR has 0-behavior for $\varphi \in [0,\varphi_C]$ and $\pi$-behavior for $\varphi \in [\varphi_C,\pi]$. (c) Critical phase $\varphi_c$ plotted as a function of $\epsilon_d$, yielding a phase diagram of the $\varphi$-controlled transition. We call $\delta \epsilon$ the width of the transition. The different samples correspond to the ones described on table \ref{table1}. More details can be found in ref \cite{Delagrange2015,Delagrange2016,DelagrangePhD}.}
	   \label{Is_scan}
\end{figure}

The modulation of the switching current $\delta I_s$ of the SQUID versus magnetic field B, proportional to the CPR, is measured for various $V_g$ and is represented on Fig.\ref{Is_scan}a together with CPRs extracted from the 0-$\pi$ transition.
	
On the edges of the diamonds, far from the transition (fig. \ref{Is_scan} (a.1)), the junction  behaves as a regular 0-junction, with a CPR proportional to $\sin(\varphi)$. In contrast, at the center of the diamond (Fig. \ref{Is_scan}.(a.6)), where $T_K$ is minimum, the CPR is $\pi$-shifted ($\delta I_s \propto \sin(\varphi + \pi)$) and has a smaller amplitude, characteristic of a $\pi$-junction. In between, the CPR is anharmonic: a distortion appears first around $\pi$ and develops as $T_K$ decreases. The CPR is composite, with 0-junction behavior around $\varphi=0$ and a $\pi$-junction behavior around $\varphi=\pi$. The transition from one part to the other is achieved by varying the superconducting phase. The global state of the system is called 0' or $\pi'$ depending on what is the dominant contribution (and where is the global minimum of the free energy, see section \ref{sec_0pitransition}).

\paragraph{Comparison with QMC calculations}
For a quantitative comparison between theory and experiment, in collaboration with D. Luitz and V. Meden, we performed a CT-INT calculation in the superconducting state ($B=0$) in diamond I to obtain the CPRs in the transition regime \cite{Delagrange2015}. 
Using the measured value of the superconducting gap $\Delta=0.17\,\mathrm{meV}$ and the previously determined parameters, the Josephson current has been computed as a function of the phase difference $\varphi$. The theoretical CPR are calculated at various $\epsilon_d$ (related to $V_g$ by $\epsilon_d=\alpha V_g$) and plotted as black lines in comparison to our experiments in fig. \ref{Is_scan} (a.2) to (a.5). Since our setup yields a switching current that is necessarily smaller than the critical current, the experimental CPRs were multiplied by a unique correction factor chosen to obtain the best agreement with the QMC results. The agreement for the shape of the CPR is excellent; however a shift of the energy level $\delta\epsilon_d=0.28~\mathrm{meV}$ of the theoretical CPRs is needed to superimpose them with the experimental ones: the QMC calculations predict a transition region centered around a smaller $\epsilon_d$ than measured experimentally \cite{Delagrange2015}. Note however that the width of this transition is very well reproduced. Interestingly, these data have been fitted by Zonda \textit{et al.} \cite{Zonda2016} with a perturbative theory: they argue that, with a charging energy of $3.4\mathrm{~meV}$ instead of $3.2$, the theory works perfectly. It seems that this discrepancy between experiment and theory would originate from a bad estimation of $U$. 

These data are thus consistent with a phase controlled level-crossing quantum transition in a single-level QD. In other words, one can control the magnetic state of the junction, doublet or singlet, with the superconducting phase.	    

\subsubsection{Universal phase diagram of the first order transition}
	
We present now a more quantitative study of the level-crossing quantum transition in the single-level regime. We call $\varphi_c$ the superconducting phase at which, at a fixed gate voltage, the system undergoes the transition from 0 to $\pi$. Theoretically, this critical phase $\varphi_c$ is defined at T=0, where the transition is expected as a jump in the supercurrent. At finite temperature the transition is rounded but, if T is small enough, $\varphi_c$ equals the phase at which the supercurrent is zero \cite{Karrasch2008,Delagrange2015}. On fig. \ref{Is_scan} (c), we show $\varphi_c(\epsilon_d)$ from experiment for different 0-$\pi$ transitions. In this figure each transition is characterized by two parameters: the value of $\epsilon_d$, called $\epsilon_t$, at which the junction changes from $\pi$ to 0 at $\varphi=\pi/2$, and the width $\delta \epsilon$ of the transition. These quantities are given in table \ref{table1} for the concerned diamonds (B, C, G and I). $\delta \epsilon$ is found to depend strongly on the parameters of the diamonds: large transition's widths correspond to ratios $T_K(\epsilon=0)/\Delta$ close to 1 (see left inset of fig. \ref{Is_scan}). To compare these eight transitions (left and right sides of four diamonds), we plot on fig. \ref{Is_scan} the critical phase $\varphi_c$ as a function of $(\epsilon_d -\epsilon_t)/\delta \epsilon$. For diamonds B, G and I, the scaled data fall on the same curve, with an arccosine dependence.

All the transitions display the same characteristic shape. We try as well to compare these data to an analytical formula obtained in the atomic limit of the Anderson impurity model with $\Delta \gg \Gamma$ and $T=0$: $\varphi_c(\epsilon)= 2 \arccos{\sqrt{g-(\epsilon/h)^2}}$: a fit of our data with this formula gives a very good agreement for $g=2.2$ and $h=0.72~\mathrm{meV}$. Note that since we are not at all in an atomic limit, these fit parameters do not correspond to the any physical one of the experiment.
	
\begin{table}
	         \renewcommand\arraystretch{1.8}
	         \begin{center}
	         \begin{tabular}{|c||c|c|c|c|c|c|c|c|c|}
	         \hline
	          Sample & \multicolumn{2}{c|}{B} & \multicolumn{2}{c|}{C} & \multicolumn{2}{c|}{G} & \multicolumn{2}{c|}{I} & J right\\
	         \hline
	         \hline
	         N  &  \multicolumn{8}{c|}{1} &  3 \\
	         \hline
	         $\Delta$  &  \multicolumn{9}{c|}{0.17 (S-NbAl)} \\
	         \hline
	         U  &  \multicolumn{2}{c|}{2.8} &  \multicolumn{2}{c|}{2.3} &  \multicolumn{2}{c|}{3.5} &  \multicolumn{2}{c|}{3.2}&  3.2 \\
	         \hline
   	         $\Gamma$  &  \multicolumn{2}{c|}{0.43} &  \multicolumn{2}{c|}{0.5} &  \multicolumn{2}{c|}{0.4} &  \multicolumn{2}{c|}{0.44}&  0.45 \\
	         \hline
	         $\Delta E$  &  \multicolumn{2}{c|}{5} &  \multicolumn{2}{c|}{4} &  \multicolumn{2}{c|}{5} &  \multicolumn{2}{c|}{4}&  4 \\
	         \hline
	         $\delta E$  &  \multicolumn{2}{c|}{0.8} &  \multicolumn{2}{c|}{0.5} &  \multicolumn{2}{c|}{0.7} &  \multicolumn{2}{c|}{0.3}& 0.3 \\
	         \hline
	         $k_B T_K$& \multicolumn{2}{c|}{$\approx$ 0.06} &  \multicolumn{2}{c|}{0.13} &  \multicolumn{2}{c|}{$\approx$ 0.03} &  \multicolumn{2}{c|}{0.05} & 0.05 \\
	         \hline
	         \hline
	         $2\delta \epsilon/U$& 0.23 & 0.43 & 0.87 & 0.96 & 0.06 & 0.06 & 0.15 & 0.2 & 0.2 \\
	         \hline
	          $2\epsilon_t/U$& 0.79 & 0.64 & 0 & 0 & 0.74 & 0.74 & 0.75 & 0.69 & 0.8 \\
	          \hline
	         
	         \end{tabular}
	         \end{center}
	             \caption{Values of occupancy N, U, $\Delta$, $\Gamma$, $\Delta E$, $\delta E$, $T_K$, $\delta \epsilon$ and $\epsilon_t$, given in meV for the investigated diamonds. For diamonds B, C, G and I, there are two transitions (0 to $\pi$ and $\pi$ to 0), with different parameters. In diamond J, only the right side of the diamond exhibits a phase dependence of the transition.}
	              \label{table1}
	         \end{table}

\subsection{Effect of the two-level regime on the 0-$\pi$ transition}
\label{CPR_2L}
	  
\begin{figure}
     \begin{center}
      \includegraphics[height=6.2cm]{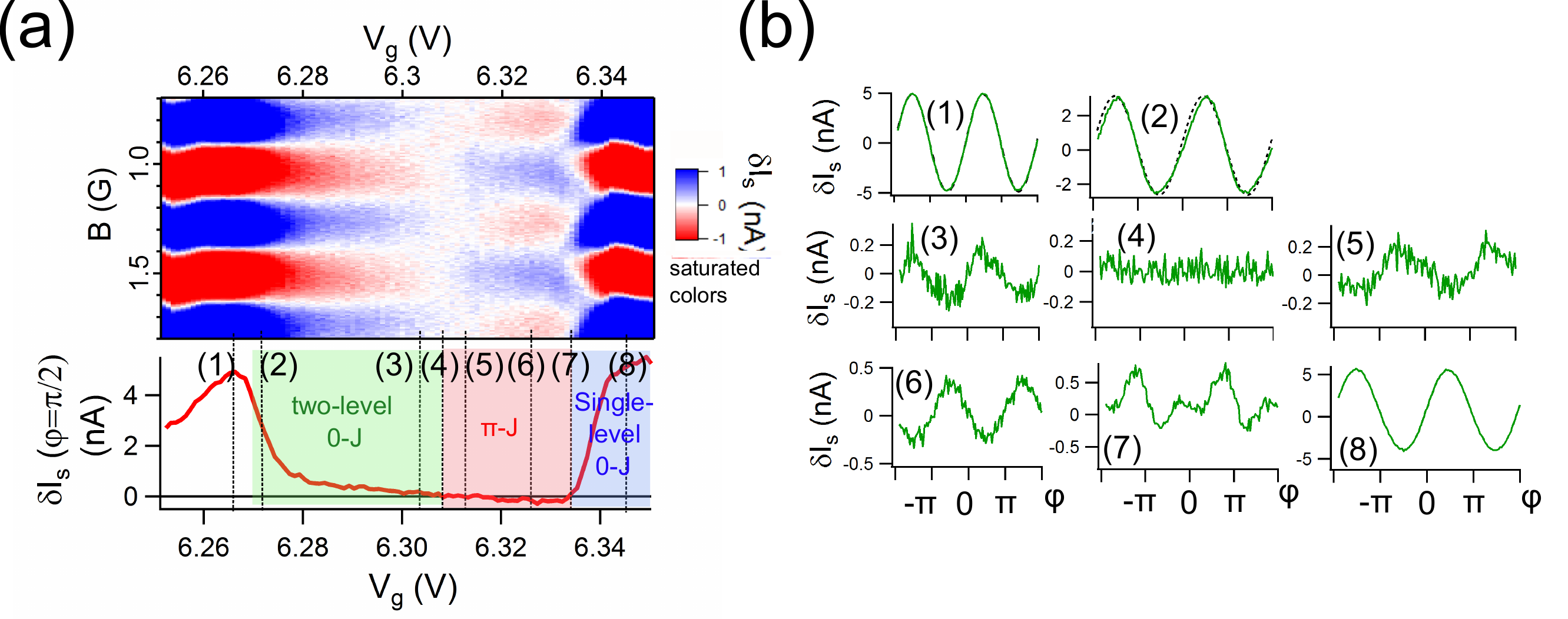}
      \end{center}
      \caption{(a) Modulation of the switching current of the SQUID versus the magnetic field, proportional to the CPR, as a function of the gate voltage $V_g$, for the diamond J. The supercurrent at the transition  being very low, the color scale is saturated. 
	  (b) Supercurrent versus the superconducting phase $\varphi$ at some particular gate voltage, indicated by the numbers on panel (a). Note the zero amplitude of the CPR (4) and the fact that the CPR (2), in the two-level 0-junction, has a stronger anharmonicity than (8), on the degeneracy point. Dashed line on (2): guide for the eyes representing a sine function, showing that the continuous line is not perfectly harmonic.More details can be found in ref \cite{Delagrange2016,DelagrangePhD}}
	 \label{detail_cpr_ML}
\end{figure}

Now we focus on the two-level 0-$\pi$ transitions, and more particularly on diamond J, corresponding to a N=3 filling factor. The modulation of $I_s$ versus the magnetic field, proportional to the CPR, is represented on Fig. \ref{detail_cpr_ML} (a) as a function of $V_g$. CPRs are also shown for some particular values of gate voltage (fig. \ref{detail_cpr_ML} (b)). 

On the right side of the diamond, close to the $N=3$ to $0$ degeneracy point (see right part of fig. \ref{2L-QD_schema}), the 0-$\pi$ transition is achieved through 0' and $\pi'$ states, similarly to the one investigated in the single-level regime (fig. \ref{detail_cpr_ML} (b) (8)). In addition, $\varphi_c=f((\epsilon -\epsilon_{t})/\delta \epsilon)$ collapses on the same arc-cosine shape as the single-level 0-$\pi$ transitions (see fig. \ref{Is_scan}).
But on the left side of the diamond, from the $N=3$ to $2$ degeneracy point to the center of the diamond (fig. \ref{detail_cpr_ML} (b) (1) to (3)) , the CPR behaves as a 0-junction. The supercurrent's amplitude decreases with $V_g$ and evolves from 0 to $\pi$ continuously. Close to the transition, around the center of the diamond (fig. \ref{detail_cpr_ML} (b) (2) to (4)), the CPR becomes slightly anharmonic. But, unlike in the other transitions, no 0' or $\pi'$ state is observed. Note as well that the current-phase relation has a larger anharmonicity in the two-level 0-junction than on the charge degeneracy points, while its amplitude is smaller (fig. \ref{detail_cpr_ML} (b) (2), to compare with fig. \ref{detail_cpr_ML} (b) (1) or (8)).
	  
	  \begin{figure}
	  	    \begin{center}
	  	    \includegraphics[height=6cm]{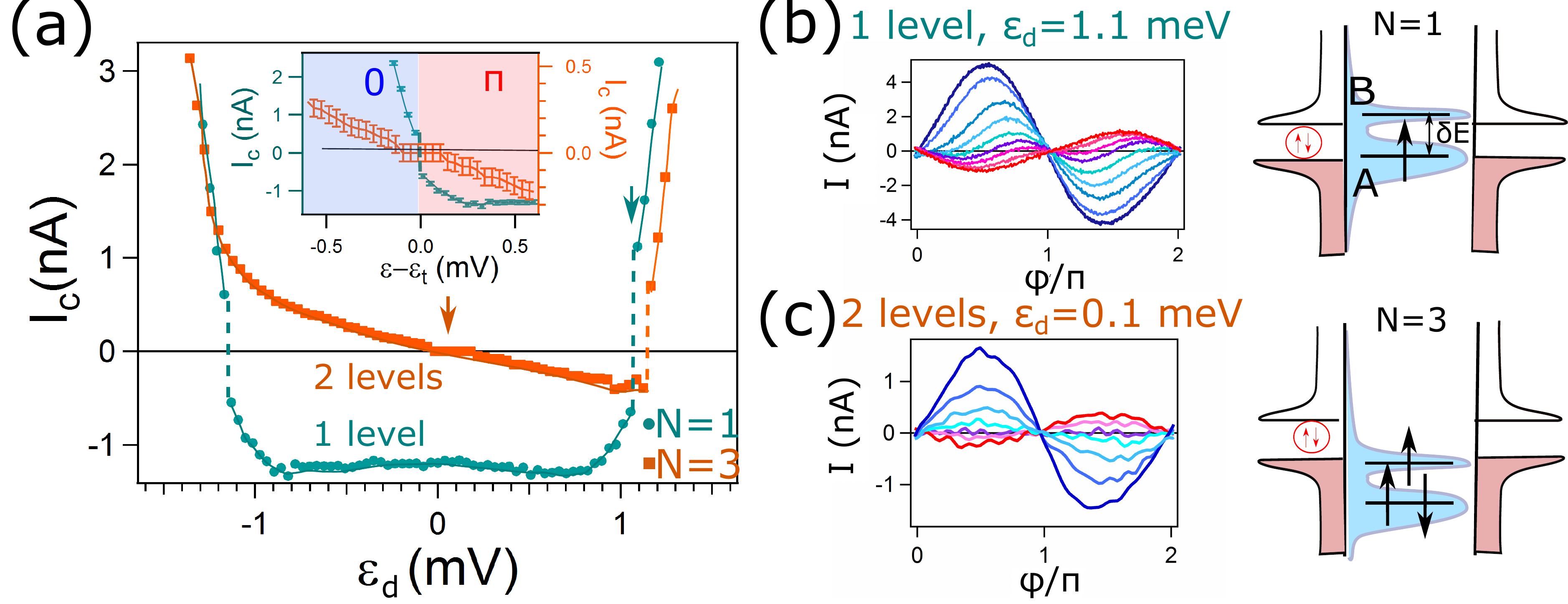}
	  	    \end{center}
	  	    \caption{(a) Critical current $I_c$, defined as the maximum amplitude of the measured switching current, as a function of the energy level $\epsilon$. $I_c$ is defined as positive for a 0-junction, and negative for a $\pi$-junction. This quantity is plotted for the two diamonds I (N=1, blue dots) and J (N=3, orange squares), respectively in the single-level and two-level regime. The dashed lines materialize discontinuities of $I_c$, specific of first order transitions. Inset: focus on two 0-$\pi$ transitions, centered around $\epsilon_t$ : at $\epsilon_d=-1.1\mathrm{~meV}$ in diamond I, and at $\epsilon_d=0.1\mathrm{~meV}$ in diamond J.
	  	    (b) Current-phase relations around the 0-$\pi$ transition in diamond I, at $\epsilon_d=1.1\mathrm{~meV}$, indicated by a green arrow. The CPR becomes anharmonic and the critical current never vanishes. (c) Current-phase relations around the 0-$\pi$ transition in diamond J, at $\epsilon_d=0.1\mathrm{~meV}$, indicated by an orange arrow. The CPR is harmonic all over the transition, and the critical current vanishes at the transition. 
	  	    (c) Schematic explanation of the symmetry breaking observed in the supercurrent between $N=1$ and $N=3$ fillings. The lowest energy level A is better coupled to the reservoirs than the highest one B (see text). More details can be found in ref \cite{Delagrange2016,DelagrangePhD}.}
	  	    \label{SL-ML}
	  	    \end{figure} 
	  	    
To go further, we consider now the critical current $I_c$. This quantity is the maximum of the CPR and is extracted here as the maximum amplitude of the modulation of the switching current, positive for a 0-junction, and negative for a $\pi$-junction. It is represented as a function of $\epsilon_d$ on fig. \ref{SL-ML} (a), for diamonds I and J, respectively in the single and two-level regime. In diamond I (N=1), the phase-dependence of the single-level transitions gives rise to discontinuities of $I_c$, which characterize a first order transition \cite{Maurand2012,Vojta2006}. For diamond J (N=3), the 0-$\pi$ transition at $\epsilon_d=0.1\mathrm{~meV}$ does not exhibit this phase-dependence, yielding a vanishing critical current at the transition. This is not anymore a first order transition, contrary to the transition at the other side of the diamond J ($\epsilon_d=1.2\mathrm{~meV}$). According to ref. \cite{Lee2010}, this kind of 0-$\pi$ transition, without intermediate state 0' or $\pi'$, is indeed possible when there is no change of the magnetic state of the system. This emphasizes that we are facing a transition between a 0-doublet state and a $\pi$-doublet (instead of a 0-singlet and a $\pi$ doublet as in the single level regime and on the right side of the diamond), specific to the two-level regime.
	  
This kind of gate dependence of the supercurrent is predicted in a carbon nanotube QD (see section \ref{Josephson_2L}), in absence of Kondo effect. The comparison of our data with ref. \cite{Yu2010a} suggests that in our experiment, the channels associated with each orbital are mixed during the transfer of Cooper pairs. This is also why we do not observe two-level induced $\pi$-junctions for even occupancies of the dot. 
	
To explain why this two-level behavior is observed for $N=3$ but not for $N=1$, we propose that the two orbital levels A and B (see fig. \ref{SL-ML}) of the CNT are slightly differently coupled to the electrodes, as in ref. \cite{Holm2008}. A detailed analysis of the gate dependence of the inelastic cotunneling peaks in the even diamond between I and J in the normal state shows indeed that $\Gamma_A\geq\Gamma_B$. Following ref. \cite{Holm2008}, we roughly evaluate $\Gamma_A-\Gamma_B\approx 0.07\mathrm{~meV}$.
	
When two quasi-degenerated levels have different widths, the supercurrent is mostly carried by the broader one. We therefore expect different behaviors for N=1 and N=3, as pointed out theoretically by Droste \textit{et al.} \cite{Droste2012}. Here, the lowest level A is more coupled to the electrodes than the highest one (B). For $N=1$, the unpaired electron occupies the level A and the level B is too poorly coupled to participate to the transfer of Cooper pairs: we are in a single-level situation, the junction is $\pi$. For $N=3$, the unpaired electron is in the poorly-coupled-level B, which thus  participates to the transport: the system is in a two-level regime. According to this interpretation, in the opposite situation of a level B better coupled than the level A, the $N=1$ diamond would exhibit the two-level physics instead of the N=3 diamond. 

There could be another reason for the electron-hole (N=1/3) symmetry to be broken in a carbon nanotube: spin-orbit coupling \cite{Makarovski2007a,Galpin2010,Droste2012}. According to Brunetti \textit{et al.} \cite{Brunetti2013}, the consequence on the supercurrent could be similar to what we observe. It is however difficult to be affirmative, since we are not able to measure the value of the SO coupling in our system. It would indeed require to study the evolution of cotunneling peaks versus the magnetic field parallel to the nanotube. 

\subsubsection{Toward a $\varphi_0$-junction?}

\begin{figure}
 	 \begin{center}
 		 \includegraphics[height=8.5cm]{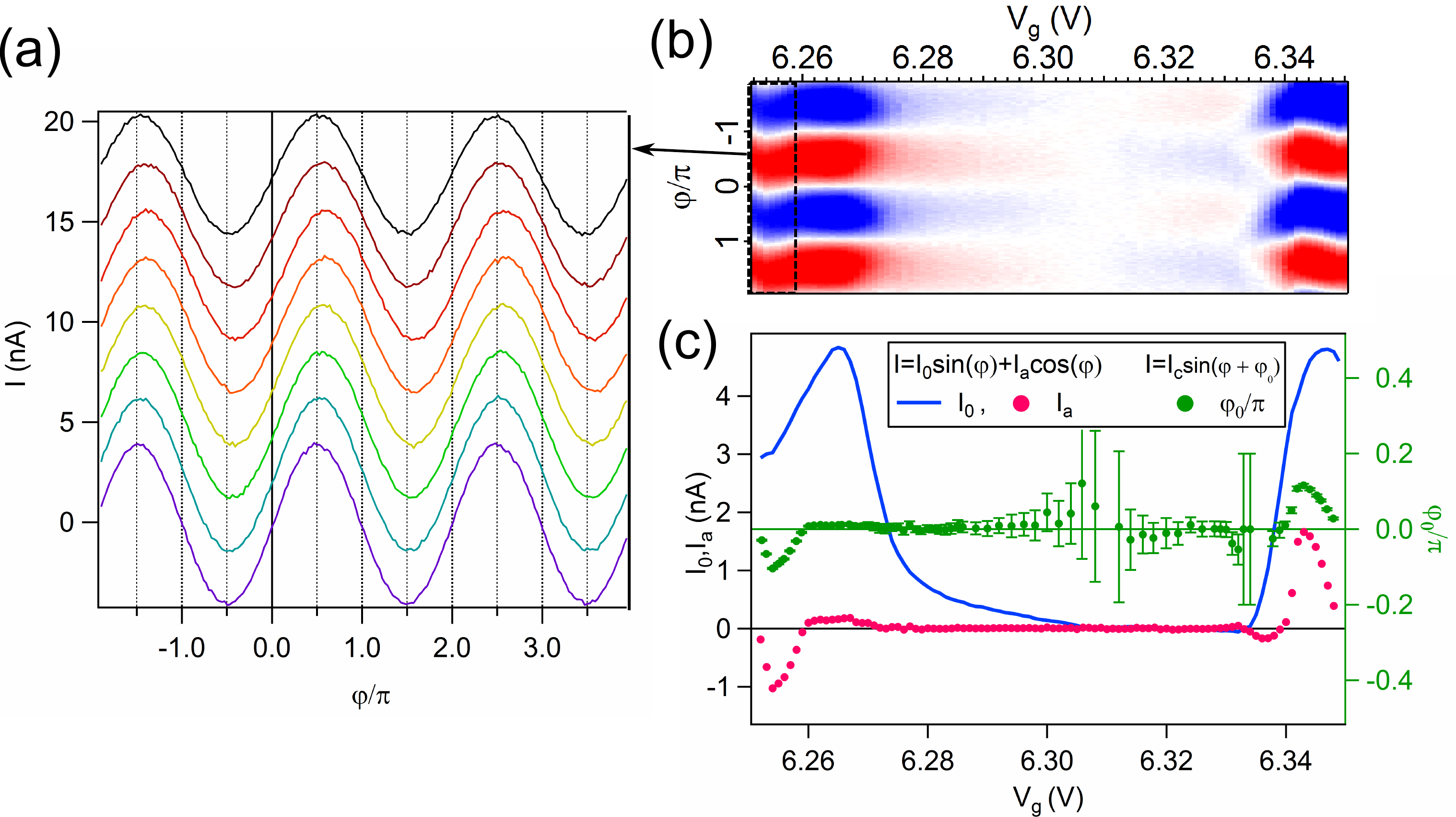}
 	\end{center}
 	\caption{(a) Current-phase relations for various values of gate voltage on the left edge of the gate voltage region (square in black on (b)). We can see that these CPRs look sinusoidal and that some of them are slightly dephased (in particular the red and orange). (c) Result of sinusoidal fits all over the diamond. In blue and pink are represented the critical current $I_0$ and the anomalous current $I_a$ from $I=I_0\sin(\varphi)+I_a\cos(\varphi)$, in green the phase shift $\varphi_0$ from $I=I_c\sin(\varphi+\varphi_0)$. Note that the error bar diverges at the 0-$\pi$ transitions: at the center where the amplitude almost vanishes, and on the right edge where the CPRs are strongly anharmonics.}
  	\label{jonction_phi}
 	 \end{figure}

On the CPR measured in the two-level regime, represented on fig. \ref{jonction_phi}, it seems that, on the edges of the gate voltage region, the CPR exhibits a small continuous phase shift. 

Until now, we have insisted on the fact that a current-phase relation is necessary an odd function of the phase: $I(-\varphi)=-I(\varphi)$ such that $I(\varphi=0)=0$. But if the time reversal symmetry (TRS) is broken in the system, it is possible to observe a non-zero current for a zero phase bias, called the anomalous current $I_a$ which can be defined as: 
\begin{equation}
I=I_0\sin(\varphi)+I_a\cos(\varphi)
\end{equation}
 (with harmonics if the CPR is non-sinusoidal). Equivalently, we can write the supercurrent as $I=I_c\sin(\varphi+\varphi_0)$, this is why such a junction is called $\varphi_0$-junction.
 
To quantify the measured effect, we represent on fig. \ref{jonction_phi} (c) the results of fit for both $\varphi_0$ and $I_0,\:I_a$ as defined above. It seems indeed that a $\phi_0$ junction develops on the edges of the diamond.
 
Is the feature on our measurements related to a breaking of the TRS? To assure this, one should prove the reproducibility of the effect, as it could be due for example to the trapping of a vortex in the coil. However, the symmetry of the effect on both edges of the diamonds makes us think about something related to Coulomb blockade.

Then the effect could be due to the presence of spin-orbit coupling in the nanotube, as well as to a small magnetic field associated to the participation of two-levels to the transport of Cooper pairs. Zazunov \textit{et al.}\cite{Zazunov2009} as well as Brunetti \textit{et al.} \cite{Brunetti2013} have shown that a substantial anomalous current could be observed in this kind of system, especially in oddly occupied diamonds, where Coulomb interactions enhance the effect. This requires further investigations, in particular measuring other samples.

\section{Conclusion}

In conclusion, we have measured in a carbon nanotube quantum dot junction the supercurrent as a function of the superconducting phase across it. We have measured this quantity in the regime where the Kondo and superconducting correlations are of the same order of magnitude and shown that the ground state of the system, singlet or doublet (corresponding respectively to 0 and $\pi$ junctions), is then controlled by the superconducting phase, giving rise to strongly anharmonic current-phase relations. We have also demonstrated that, if a second energy level participate in the transport of Cooper pairs, the 0-$\pi$ transition is not anymore a first order one, as it is the case when only one level is involved.

\section*{Acknowledgments}

The authors acknowledge S. Gu\'eron, J. Basset, P. Simon, A. Murani, F. Pistolesi, S. Florens and M. Filippone for discussions and technical help from S. Autier-Laurent. 
This work was supported by the French program ANR MASH (ANR-12-BS04-0016), DYMESYS (ANR 2011-IS04-001-01) and DIRACFORMAG (ANR-14-CE32-0003).

\end{document}